\documentclass[twocolumn,prb]{revtex4}

\usepackage{graphicx}
\usepackage{subfigure}
\usepackage{textcomp}
\usepackage{amsmath}

\newcommand{\nv}{NV}
\newcommand{\tat} {$^3A_2$}
\newcommand{\ct}{$^{13}$C}
\newcommand{\bra}[1] {\langle #1 |}
\newcommand{\ket}[1] {| #1 \rangle}

\newcommand{\me}[3] {\bra{#1} #2 \ket{#3}}
\newcommand{\mspin}[1]{$m_s =$ #1}

\begin{document}
	
\title{Temperature dependence of the $^{13}$C hyperfine structure of the negatively-charged nitrogen-vacancy center in diamond}

\author{M.S.J. Barson$^1$}\author{P.M. Reddy$^1$}\author{S. Yang$^{2,3}$}\author{N.B. Manson$^1$}\author{J. Wrachtrup$^2$}\author{M.W. Doherty$^1$}

\address{$^1$Laser Physics Centre, Research School of Physics and Engineering, Australian National University, 2601, Australia}
\address{$^2$3.\,Physikalisches Institut, Universit{\"a}t Stuttgart, Pfaffenwaldring 57, Stuttgart, 70569, Germany}
\address{$^3$Department of Physics, The Chinese University of Hong Kong, Shatin, New Territories, Hong Kong, China}

\begin{abstract}
	The nitrogen-vacancy (NV) center is a well utilized system for quantum technology, in particular quantum sensing and microscopy. Fully employing the NV center's capabilities for metrology requires a strong understanding of the behaviour of the NV center with respect to changing temperature. Here, we probe the NV electronic spin density as the surrounding crystal temperature changes from 10 K to 700 K by examining its \ct\, hyperfine interactions. These results are corroborated with \textit{ab initio} calculations and demonstrate that the change in hyperfine interaction is small and dominated by a change in the hybridization of the orbitals constituting the spin density. Thus indicating that the defect and local crystal geometry is returning towards an undistorted structure at higher temperature.
\end{abstract}

\maketitle

\section{Introduction}

The negatively charged nitrogen-vacancy (NV) center in diamond is a leading system in quantum technology. Due to its atom-like size, bright fluorescence, ability to optical initialize and read-out its electron spin, and long coherence time in ambient conditions, the NV center has been rapidly adopted for nanoscale quantum sensing/microscopy and quantum information processing (QIP). The quantum sensing applications of the NV center take advantage of the susceptibilities of the ground-state spin-resonances to magnetic fields \cite{loubser1977optical,balasubramanian2008nanoscale,maze2008nanoscale,taylor2008high}, electric fields \cite{dolde2011electric,dolde2014charge}, temperature \cite{Acosta2010,chen2011temperature,neumann2013high,toyli2012measurement,toyli2013fluorescence} and strain \cite{nanomechanical_sensing_barson,teissier2014strain,ovartchaiyapong2014dynamic}. Precision quantum sensing and high-fidelity QIP operations require these susceptibilities to be well characterized and understood. The magnetic, electric and strain responses have been well characterized and understood. However, there remains contention regarding the origin of the temperature susceptibility. Initial models considering only strain from thermal expansion failed to describe the observed behaviour \cite{chen2011temperature, Acosta2010, toyli2012measurement}. To rectify the unexplained temperature dependence, Doherty \textit{et al} \cite{PhysRevB.90.041201} added a quadratic spin-phonon interaction to phenomenologically explain the thermal dependence. To fully accept this model, further first principles modelling and experimental validation is required. Here, we use the hyperfine interaction to probe the electronic behaviour during thermal expansion via experiment and first principles modelling.

\begin{figure}
	\centering
	\subfigure[]{\includegraphics[width=0.2\textwidth]{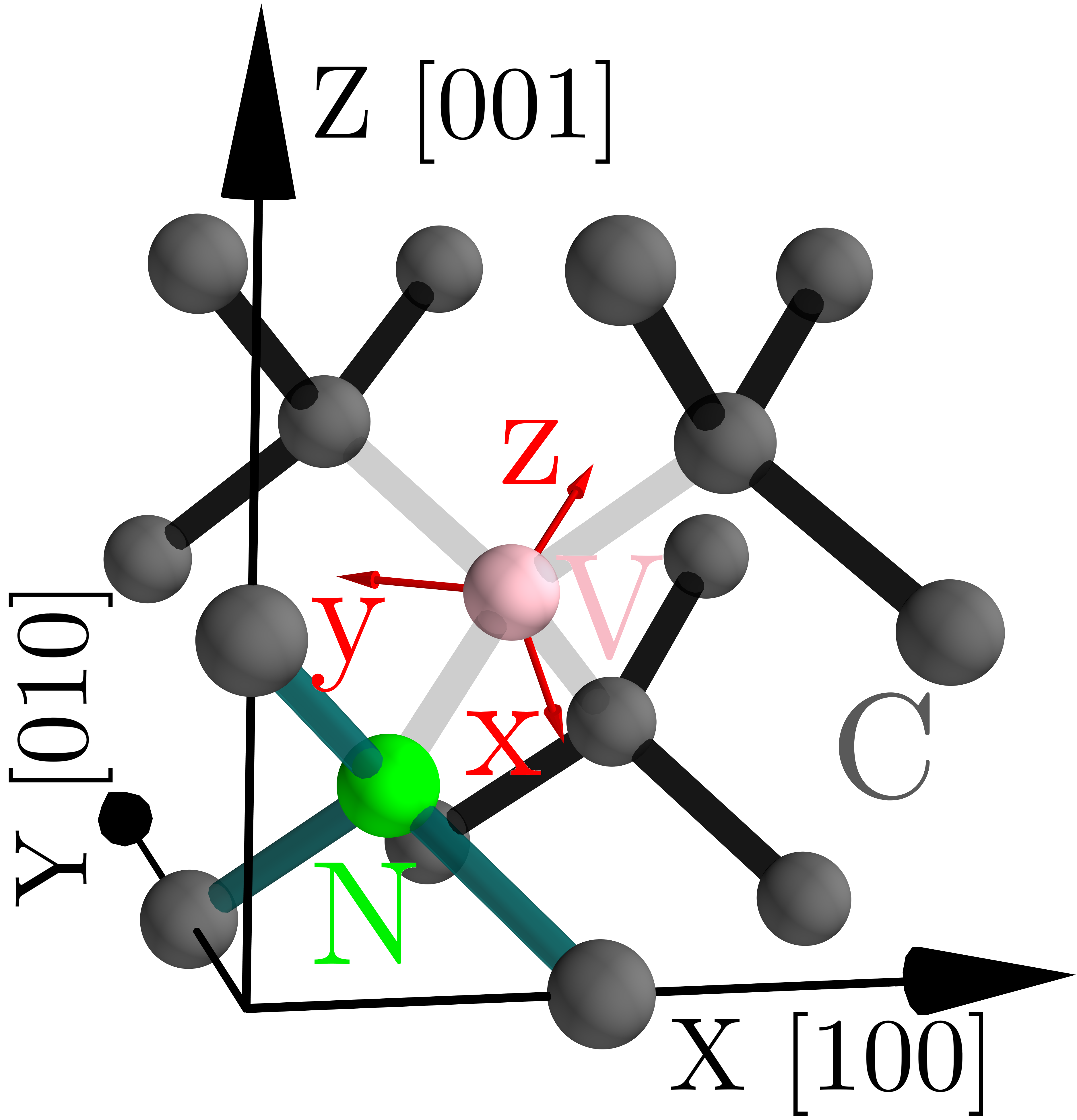}\label{fig-NV-structure}}
	\subfigure[]{\includegraphics[width=0.2\textwidth]{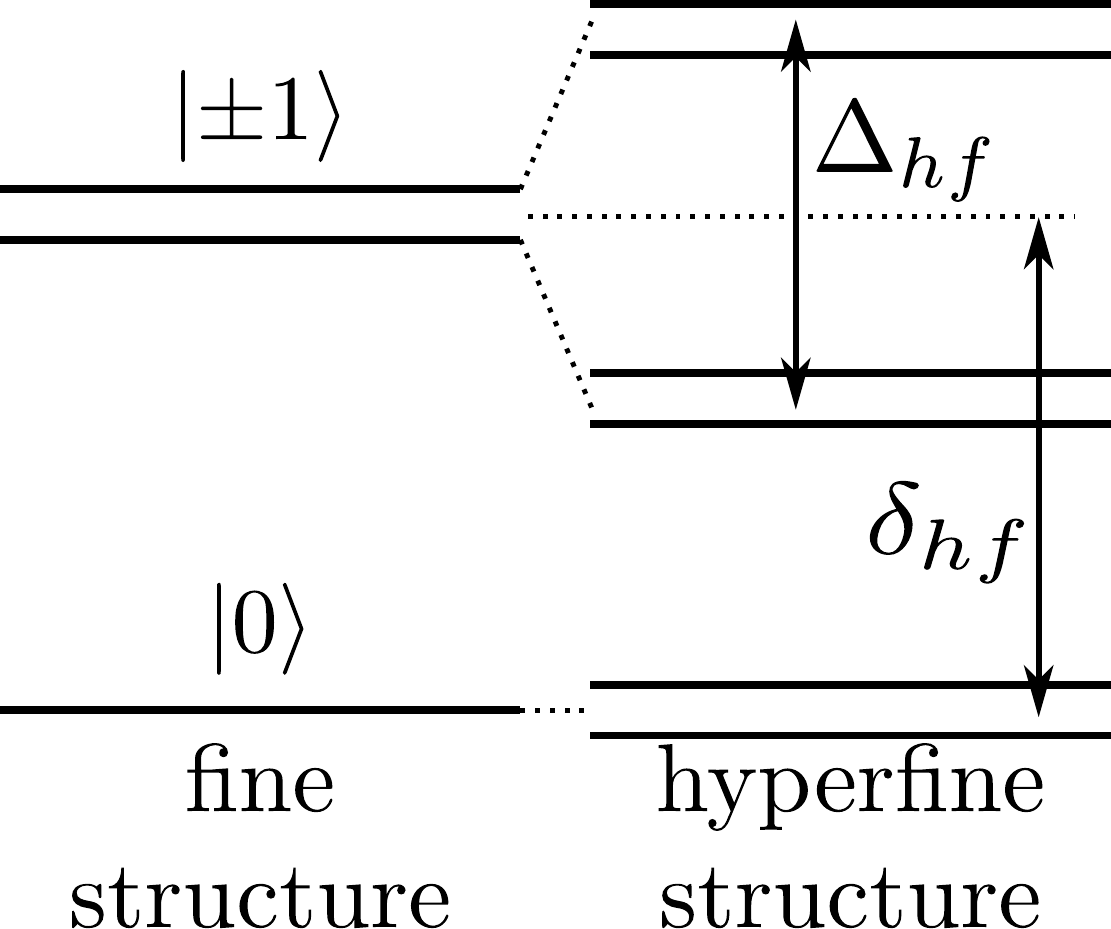}\label{fig-level_diagram}}
	\subfigure[]{\includegraphics[width=0.3\textwidth]{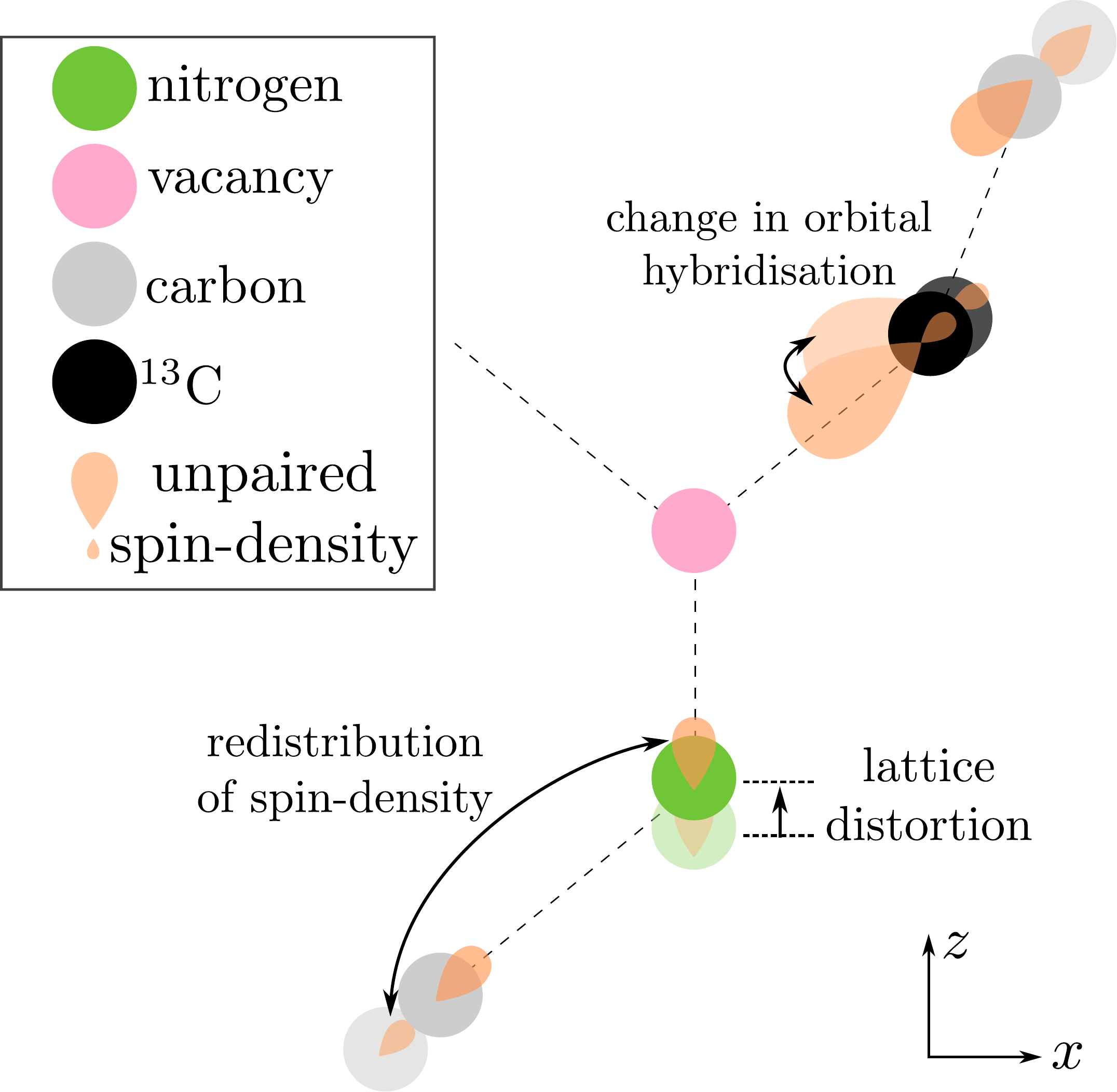}\label{fig-sp3-bond-etc}}
	\caption{\textbf{(a)} Unit cell with NV coords $(x,y,z)$ and crystallographic coordinates $(X,Y,Z)$. \textbf{(b)} Energy levels with the \ct\, hyperfine interaction with spin-spin splitting $\delta_{hf}$ and hyperfine splitting $\Delta_{hf}$. Due to the non-diagonal components of the hyperfine interaction, for zero or small applied magnetic fields the hyperfine levels are a mixture of spin levels, without simply defined quantum numbers. \textbf{(c)} Depiction of processes occurring during thermal expansion; lattice distortion, redistribution of spin density between atoms and rehybridization (reorientation) of the orbitals comprising the spin density.e.}
\end{figure}

During thermal expansion/contraction or mechanical stress, the nuclei surrounding the NV change position, causing the NV center's electron orbitals to move and change \cite{doherty2014electronic, nanomechanical_sensing_barson}. The effect of this change on the hyperfine interaction is included in two ways: (1) the orbital hybridization (or bond-angle) of the atomic orbitals constituting the unpaired spin density changes, and (2) the spin-density of the electron at the location of the nuclear spin changes. These processes are depicted in figure \ref{fig-sp3-bond-etc}.  Precise information of these processes will improve our fundamental understanding of the \nv\, electronic orbitals and how they change with temperature.


The hyperfine interaction is due to two contributions: (1) a dipolar interaction between the nuclear spin and the local distribution of electron spin density and (2) the Fermi contact interaction of the electron spin density at the nuclear spin \cite{loubser1978electron}. Both of these contributions are strongly dependent on the relative positions of the nuclear and electron orbitals. As such, the hyperfine interaction can be used as a atomscopic probe of the effect of thermal expansion on the \nv\, electron orbitals.

This investigation first outlines the hyperfine interaction theoretically and its relation to the \nv\, electronic wavefunction. Experimental tests are then performed and first principles modelling is used to provide further insight into the results. The key outcomes are that the change in hyperfine interaction is negligible and the dominant change in the electron spin density is a variation in atomic orbital hybridization.

\section{Theoretical Details}

The \nv\, center is an axially symmetric defect with three-fold $C_{3v}$ symmetry (figure \ref{fig-NV-structure}). The electronic ground-state (\tat) consists of an spin-triplet with a zero-field splitting of $D \approx2.87$ GHz (at room temperature) between the \mspin{0} and \mspin{$\pm1$} spin levels (figure \ref{fig-level_diagram}). The hyperfine interaction of the \nv\, ground-state with a nearby nucleus can be described by the following spin-Hamiltonian \cite{doherty2012theory},
\begin{equation}
{H}_{hf} = \mathbf{S}\cdot\mathbf{D}\cdot\mathbf{S} + \gamma_e \mathbf{S}\cdot\mathbf{B} + \mathbf{S}\cdot\mathbf{A}\cdot\mathbf{I},
\label{eqn-hyperfine_hamiltonian}
\end{equation}
where $\mathbf{S}$ are the spin-operators; $\mathbf{D}$ is the zero-field (or spin-spin) tensor; $\gamma_e$ is the electron gyromagnetic ratio; $\mathbf{B}$ is the applied magnetic field; $\mathbf{A}$ is the hyperfine tensor; and $\mathbf{I}$ are the nuclear spin operators. The temperature dependence of the zero-field splitting has been measured previously and can be described by a power series $D(T)\approx \sum d_i T^i $ \cite{Acosta2010, PhysRevB.90.041201, chen2011temperature, toyli2012measurement}. Note that the nuclear Zeeman interaction is negligible for the magnetic fields considered here ($<$ 100 G). In the diagonalized form there are only two unique components to $\mathbf{A}$, an axial ($A_\parallel$) and a non-axial ($A_\perp$) component.
\begin{align}
\mathbf{A}_\text{diag} = \left( 
\begin{array}{ccc}
A_\perp & 0 & 0 \\
0 & A_\perp & 0 \\
0 & 0 & A_\parallel
\end{array} \right).
\end{align}
These components are given with respect to a reference frame that is not the NV coordinate system \cite{felton2009hyperfine,He1993paramagnetic-ii,loubser1977optical,loubser1978electron}. As such, a rotation is required to transform the hyperfine tensor to the NV coordinate system $\mathbf{A} = R\cdot\mathbf{A}_\text{diag}\cdot R^T$, see appendix \ref{section-rotations} for details. The hyperfine parameters for a first shell \ct\, are $A_\parallel=199.7(2)$ MHz and $A_\perp=120.3(2)$ MHz \cite{felton2009hyperfine}. Solving the Hamiltonian either numerically or by approximation (see appendix \ref{section-rotations}) gives the \ct\, zero-field hyperfine splitting $\Delta_{hf} \approx 127$ MHz and average resonance $\delta_{hf} = 2876$ MHz. Due to the non-diagonal terms in $A$ and the large size of the hyperfine perturbation relative to the zero-field splitting, there is a small amount of $D$ dependence in $\Delta_{hf}$ and a small amount of hyperfine dependence in $\delta_{hf}$. The large splitting $\Delta_{hf}$ enables the easy identification of \nv\, centers with a first shell \ct\, upon examination of the optically detected magnetic resonance (ODMR) spectra. 

The hyperfine components of $\mathbf{A}$ can be re-described by the Fermi contact term $f$ and the dipolar term $d$ \cite{loubser1978electron},
\begin{align}
\begin{split}
A_\parallel &= f + 2d \\
A_\perp &= f - d.
\end{split} \label{eqn-A-f-d}
\end{align}

The contact term is spatially isotropic and satisfied by a spherically distributed $s-$orbital. There is no dipolar contribution from an $s-$ orbital, as such, some $p-$ (or higher) orbital character is required. Using these arguments, the contribution to the total molecular orbitals (MOs) from a atomic orbital at the nuclear spin is a hybrid orbital $\psi$ which is a linear combination of $s-$ ($\phi_s$) and $p-$orbitals ($\phi_p$) \cite{loubser1978electron},
\begin{align}
\psi &= c_s\phi_s + c_p\phi_p \\
1 &= \left|c_s\right|^2 + \left|c_p\right|^2, \nonumber
\end{align}
where the Fermi contact $f$ and dipolar terms $d$ can be described as \cite{He1993paramagnetic-ii},
\begin{align}
f &= \frac{8\pi}{3} \frac{\mu_0}{4\pi}g_e\mu_Bg_n\mu_n\left|c_s\right|^2\eta\left|\phi_s(0)\right|^2 \label{eqn-hyperfine-f} \nonumber \\
&= 3777\times (1-\left|c_p\right|^2) \eta \;\text{MHz}, \\
d &= \frac{2}{5} \frac{\mu_0}{4\pi}g_e\mu_bg_n\mu_n\left|c_p\right|^2\eta\me{\phi_p}{\frac{1}{r^3}}{\phi_p} \label{eqn-hyperfine-d} \nonumber \\
&= 107.4\times  \left|c_p\right|^2 \eta \;\text{MHz},
\end{align}
where $\left|\phi_s(0)\right|^2$ is the probability per unit volume of $s-$orbital at the location of the \ct\, nucleus and $\me{\phi_p}{\frac{1}{r^3}}{\phi_p}$ is the average $1/r^3$ value of the $\phi_p$ orbital. $\eta$ is the electron spin-density at the nucleus, $g_n$ is the \ct\, nuclear $g-$factor and $\mu_N$ is the nuclear magneton. The atomic parameters $\left|\phi_s(0)\right|^2$ and   $\me{\phi_p}{\frac{1}{r^3}}{\phi_p}$ are numerically determined and available in published tables \cite{ayscoughEPR1967, MORTON1978577}. These relationships show that measuring $A_\parallel$ and $A_\perp$ to determine $f$ and $d$ allows for the determination of the temperature dependent changes to the \nv\, orbital hybridization $\left|c_s\right|^2$ / $\left|c_p\right|^2$ and the spin-density $\eta$.

\section{Experimental details and results}
\label{section-exp-details}

The measurements included low temperature ODMR of measurements of single NV centers and high temperature measurements of an ensemble of NV centers. For high temperature ODMR detection a lock-in amplifier (Stanford SR830) was used with a time-constant of 1 ms. The microwave signal was amplitude modulated at 100\% depth at a modulation frequency of 1.7 kHz (R\&S SMIQ03B). The increased sensitivity from the lock-in amplifier allowed for \ct\, ODMR resonances to be easily measured above background noise, (figure \ref{fig-3}(a)), even for the relatively low ${\sim} 1\%$ natural abundance of \ct\, expected in the ensemble. The optical transition of NV$^-$ was excited using a 500 mW 532 nm CW laser into a Nikon LU Plan $100\times$/0.8 WD 3.5 mm objective lens. The microwave ground state transition was excited using a simple coaxial circular loop short ($R \sim 3$ mm) and an amplifier providing 16 mW of microwaves into the short. The sample was mounted onto a home-made iron hotplate with a vacuum chuck (500 Torr) ensuring good thermal contact on the sample and a feedback loop controlling the temperature.

For low temperature single site \ct\, measurements, a liquid helium cold finger continuous flow cryostat was used with a scanning confocal microscope. A 0.9 NA Nikon LU Plan Fluor air objective was mounted on a 3 axis piezo scanning stage within the vacuum space of the cryostat. Microwaves were provided to the sample via a 25 $\mu$m wire soldered across the sample. Since single NV centers with an adjacent first shell \ct\, were required, an alpha-numerical grid and marker system was scribed into the diamond using focussed ion beam (FIB). Due to the low probability of finding a single NV center with a \ct\, in the first shell, a search algorithm was employed that recorded the position of bright spots and performed CW ODMR at each bright spot to establish if that bright spot was an NV center with a first shell \ct. For each temperature the desired NV center was re-found using the marking system due to thermal expansion of the cold-finger.

\begin{figure}
	\centering
	\includegraphics[width=0.40\textwidth]{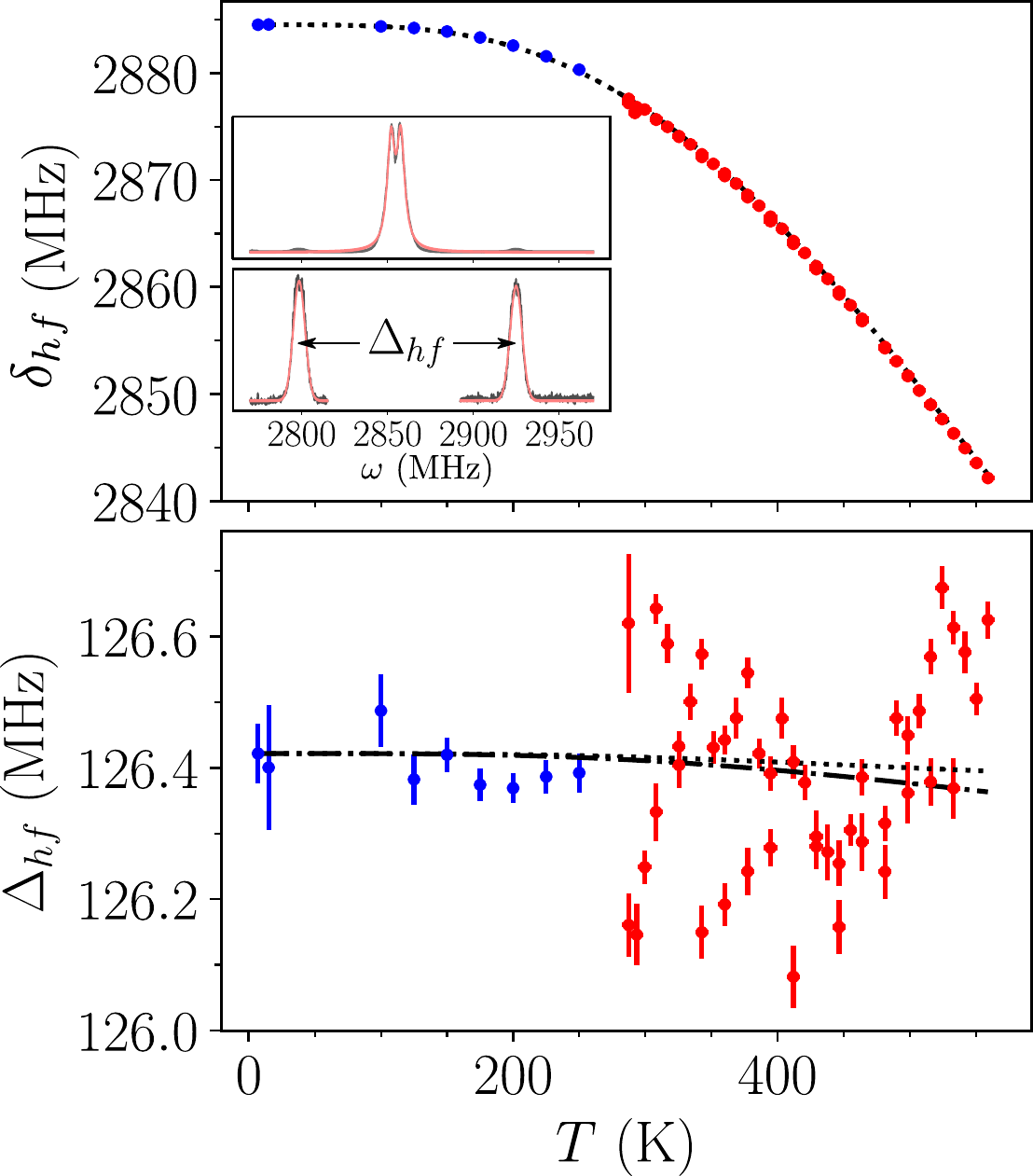}
	\caption{\textbf{(a)} Change in mean position ($\delta_{hf}$) of spin-resonances with temperature, the dashed line is from the Hamiltonian using the theroetical description for $D(T)$ (see appendix \ref{section-DT}) and the \textit{ab initio} determined hyperfine terms. Inset: Example \ct\, ODMR spectra from ensemble at T = 180\textdegree C, colored lines are Gaussian fits. The upper/lower insets are from lower/higher gain settings on the lock-in amplifier. The \ct\, resonance are just visible in the low gain spectra. The central resonance (omitted) saturates the amplifier in the high gain spectra. \textbf{(b)} Change in difference of spin-resonances ($\Delta_{hf}$) with temperature, dashed-dot line represents Hamiltonian (shifted in energy to match data) with \textit{ab initio} temperature dependent hyperfine parameters and the dotted line is with constant hyperfine parameters and $D(T)$.}
	\label{fig-2}

\end{figure}

Measurements were obtained using CW ODMR spectra for both single and ensembles of \nv\, centers. In these measurements, fluorescence is continually measured as a frequency sweep of microwaves is applied directly to the sample. An example of the \ct\, spectra can be seen in figure \ref{fig-2}(a). Repeating for a range of temperatures, comparisons of the peak frequencies of the two hyperfine resonances ($\omega_1, \omega_2$) with the mean frequency $\delta_{hf} = (\omega_1+ \omega_2)/2$ and difference in frequency $\Delta_{hf} = (\omega_1-\omega_2)$ can be made. 

As described in the last section and equations (\ref{eqn-D_hf}-\ref{eqn-Delta_hf}) in appendix \ref{section-rotations}, there is a component of $D(T)$ present in the separation of the spin-resonances, as such a small temperature dependence in $\Delta_{hf}$ inherited from $D(T)$ is expected. Despite this, the difference of the two resonances ($\Delta_{hf}$) shown in figure \ref{fig-2}(b) show no obvious change within experimental error, this indicates there is no measurable change in the hyperfine interaction. As expected, the mean frequency of the two hyperfine resonances follows the well known $D$ temperature shift, shown in figure \ref{fig-2}(a). There is no observable change in $\Delta_{hf}$ that can be attributed to a change in the hyperfine parameters $A_\parallel$ or $A_\perp$. 
The high-temperature data shows a large amount of scatter in the data points. The error bars are only indicative of the error in the fit and measurable experimental parameters, as such, there must be an larger unmeasured and unknown random error in the measurement. This is probably due to some thermal dependent fluctuations in magnetization of the iron hotplate or stray magnetism from the heater element circuitry, see appendix \ref{section-appendix-ensemble-odmr-data}. Note that the error analysis on the fitting was performed by a Monte-Carlo method, in which the raw spectra was repeatedly modulated by normal distributed random noise and re-fitted. The amplitude of this normally distributed random noise is equal to the standard deviation of the residuals from the initial fit.

\section{\textit{ab initio} calculations}

Since the temperature dependence of the hyperfine parameters was not able to be measured, further \textit{ab initio} investigation were pursued. We performed calculations using the Vienna Ab initio Simulation Package (VASP) \cite{vasp1996Kresse}. The calculations used a supercell containing 512 atoms, a plane-wave cut-off energy of 600 eV, Gamma point sampling and the Perdew-Burke-Ernzhof (PBE) functional \cite{PBEBurke1996}. The hyperfine tensor was evaluated in the \nv\, ground state for the nearest-neighbour carbon atoms to the vacancy using the inbuilt hyperfine routine of VASP.

\begin{figure}
	\centering
	\includegraphics[width=0.48\textwidth]{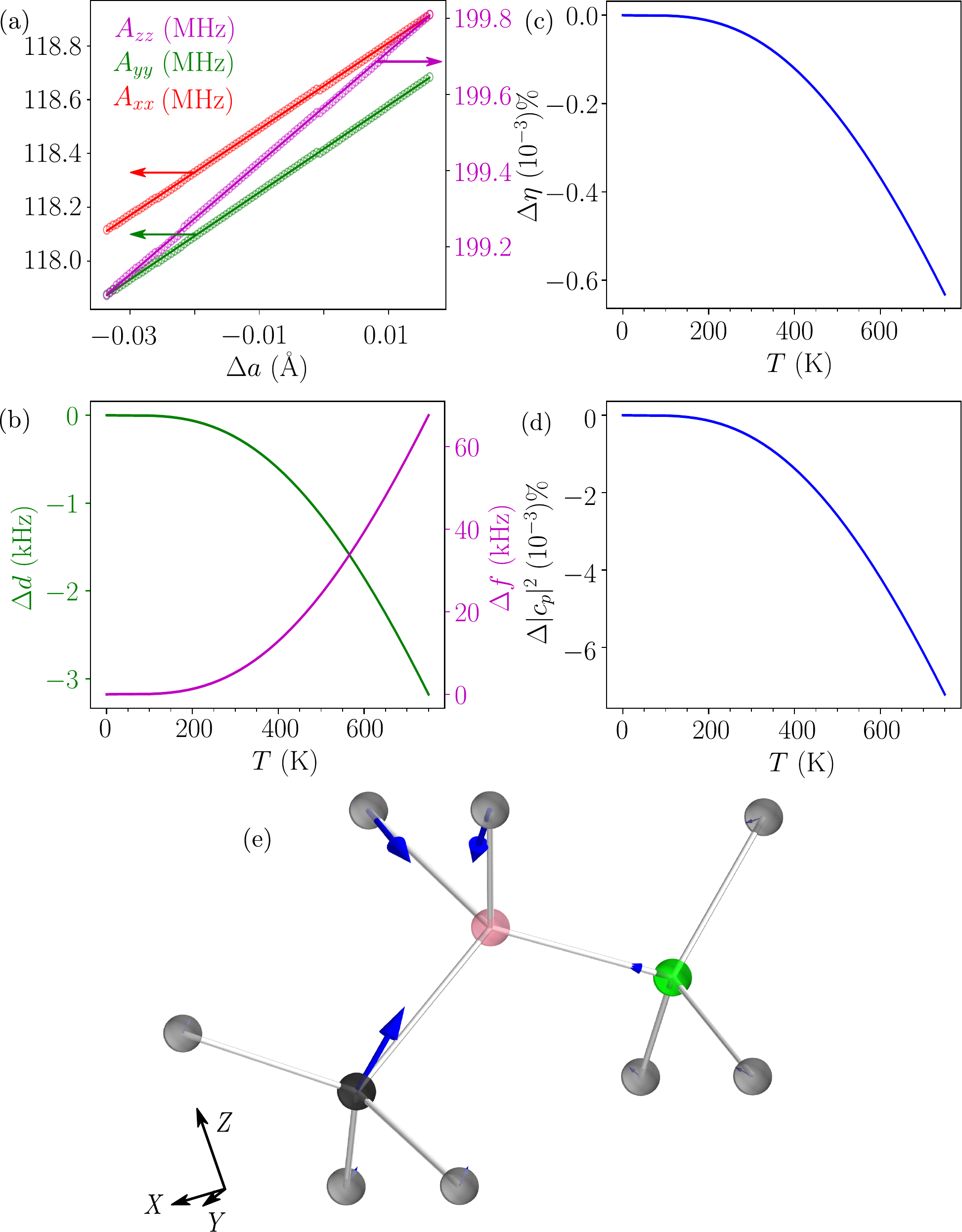}
	\caption{\textbf{(a)} \textit{Ab initio} results of hyperfine parameters (points) and linear fits (lines) due to changes in lattice constant $a$. \textbf{(b)} Changes to the dipolar ($d$) and Fermi ($f$) components of the hyperfine interaction with respect to changes in the  temperature $T$. \textbf{(c-d)} Changes to the spin-density $\eta$ and component of $p-$ orbital $\left| c_p\right|^2$. \textbf{(e)} \textit{Ab initio} results of the positions of atoms at zero temperature. The arrows indicate the shift of the atoms for increasing temperature. The arrows representing displacement are normalized in size, as such, the carbons nearest to the vacancy experience the greatest displacement. The carbon atoms are grey, the vacancy is pink, the nitrogen is green and the \ct\, atom is black. The axes describe the crystallographic coordinates $X,Y,Z = [100],[010],[100]$ and the NV axis is along the $[111]$ direction.}
	\label{fig-3}
\end{figure}

Temperature was simulated via the thermal expansion of the diamond lattice. This was implemented in the calculations by varying the lattice constant of the supercell. For each lattice constant, the atomic geometry was allowed to relax prior to the evaluation of the hyperfine tensor. A linear fit to each hyperfine tensor component was performed to obtain the components as continuous functions of lattice constant (figure \ref{fig-3}(a)). These were then converted to functions of temperature by employing the X-ray crystallography data \cite{PhysRevB.65.092102, reeber1996thermal} for the lattice constant as a function of temperature. Strong agreement with the experimental determined hyperfine parameters was obtained with the parameters with $A_\parallel \approx 199.6$ MHz and $A_\perp \approx 119.5$ MHz at zero temperature. These results are within 1$\sigma$ and 4$\sigma$ respectively of the experimental values published by Felton \textit{et al} \cite{felton2009hyperfine}.

Figures \ref{fig-3}(c-d) show the behaviour of $\left| c_p\right|^2$ and $\eta$ due to changes in temperature. These results show that the spin-density reduces slightly and the orbitals are becoming more $s-$ type and less $p-$ type. Previous \textit{ab initio} studies \cite{larsson2008electronic} have shown that the NV center self-distorts away from tetrahedral coordination at low temperature. The nearest-neighbour carbon atoms move towards being in-plane with the next-to-nearest carbon atoms. As a result, the bonds between the nearest-neighbour and next-to-nearest carbon atoms become more like $sp^2$ bonds. The dangling bond that constitutes the spin density at the nearest-neighbour carbon atoms thus becomes more like a pure $p-$orbital. These \textit{ab initio} results show that, for increasing temperature, the dangling orbital is returning towards $sp^3$, which in turn implies that the local crystal is returning towards tetrahedral structure. This is shown in figure \ref{fig-3}(e) by the large displacements of the nearest-neighbour carbons towards the vacancy.

The results also show that the spin density associated with the nearest-neighbour carbon atoms is decreasing with expansion. This must be the consequence of the spin density being distributed over more carbon atoms with expansion. This can be understood as the consequence of the confining electrostatic potential of the defect becoming shallower as the lattice expands, thus allowing the electrons to spread over more distant sites.

The overall increase in magnitude of the hyperfine parameters with expansion is consistent with this picture because the small decrease in spin density at the nearest-neighbour carbon atoms is more than compensated by the increase in $s-$orbital contribution to the spin density at those atoms. This increase in $s-$orbital increases the Fermi contact contribution, which is the dominant contribution to the parameters.

\section{Conclusion}

High and low temperature spectroscopy of the \nv --\ct\, hyperfine interaction was performed to obtain the changes in the hyperfine parameters due to changing temperature. It was experimentally found that the change in the hyperfine parameters due to changes in temperature was not resolved by our measurement. It would be possible to perform ENDOR Raman-heterodyne \cite{manson1990raman} or pulsed NMR measurements in the future to improve the accuracy of the hyperfine terms. \textit{Ab initio} simulations corroborated this evidence and found that the spin-density doesn't change significantly and there is a slightly larger change in orbital hybridization such that the bonds change hybridization towards to perfect $sp^3$ bonding at higher temperature. This seems to fit with the temperature model proposed by Doherty \textit{et al} \cite{PhysRevB.90.041201}. That is, the change in spin-density is small so that the change of spin-spin interaction due to thermal expansion predominately comes from the fact that atoms are further apart only. Ultimately, this means that the change in spin-resonances from thermal expansion is small, supporting the addition of the extra quadratic spin-phonon interaction. 

This is the first ever attempt to atomscopically measure the spin-density and hybridization ratio of the NV center's orbitals due to deformations of the diamond lattice. The small magnitude of these changes have positive implications for quantum information applications using a \ct\, spin cluster \cite{Neumann1326,dutt2007quantum}. As the resonances will not shift due to temperature fluctuations reducing the experimental requirements on thermal control or heat management. These results also rule out any benefit of using the \ct\, spin temperature susceptibility directly for thermometry applications, despite the increased coherence time of nuclear spins.

\section{Acknowledgements}

This work was supported by funding from the Australian Research Council (DP140103862, DE170100169). MSJB acknowledges support from the DAAD-GO8 joint research cooperative scheme and the Robert and Helen Crompton award. This research was undertaken with the assistance of resources and services from the National Computational Infrastructure (NCI), which is supported by the Australian Government. JW acknowledges funding from the DFG and the EU via ERC grant SMeL and ASTRIQS as well as QIA.

\appendix

\section{Hyperfine tensor rotation}
\label{section-rotations}
The most widely accepted hyperfine parameters are given by Felton \textit{et al} \cite{felton2009hyperfine} where, $A_\parallel$ and $A_\perp$ are 199.7(2) MHz and 120.3(2) MHz respectively. However, these values are given in reference to the nuclear coordinate system at an angle ($\theta$, $\varphi$) = ($125.26^\circ$, $45^\circ$) from the crystallographic ([100],[010],[001]) coordinate system. This coordinate system is aligned along $[11\bar{1}]$, which is at an angle $\theta = \arccos\left(\frac{-1}{3}\right)$ from $[\bar{1}\bar{1}\bar{1}]$. To convert to the NV spin coordinate system (where $z$ is parallel to [111],[$\bar{1}\bar{1}\bar{1}$]) a rotation operation needs to be applied to the hyperfine tensor. Ignoring the angle $\varphi$ as the system is axially symmetric about [111], the rotation matrix is,

\begin{align}
R=\left(
\begin{array}{ccc}
-\frac{1}{3} & 0 & \frac{2 \sqrt{2}}{3} \\
0 & 1 & 0 \\
-\frac{2\sqrt{2}}{3}  & 0 & -\frac{1}{3} \\
\end{array}
\right).
\end{align}
Using this rotation gives the non-diagonal hyperfine tensor in the NV coordinate frame $\mathbf{A} = R\cdot \mathbf{A}_\text{diag} \cdot R^T$

Apply second-order degenerate perturbation theory the average spin-resonance frequency is
\begin{align}
	\delta_{hf} &\approx D + \frac{4A_\parallel^2+5A_\perp^2}{12 D}, \label{eqn-D_hf} \\
	\delta_{hf} &\approx 2876 ~~\text{MHz} \nonumber 
\end{align}
and the difference in spin-resonance frequency is
\begin{align}
	\Delta_{hf} &\approx \frac{1}{18D}\left[36\left(A_\parallel^2+8A_\perp^2\right)D^2 \right. \nonumber \\
	&\qquad\left. + 324 A_\parallel A_\perp D + \left(4 A_\parallel^2+ A_\perp^2\right)^2\right]^\frac{1}{2} \label{eqn-Delta_hf} \\
	\Delta_{hf} &\approx 127.7 ~~\text{MHz}. \nonumber
\end{align}
For all of the analysis in the this report, the numerical solutions to the Hamiltonian were used, this approximation is used only to demonstrate the behaviour of the spin-resonances.

\section{Full temperature range fit for $D(T)$}
\label{section-DT}

\begin{figure}[h]
	\centering
	\includegraphics[width=0.45\textwidth]{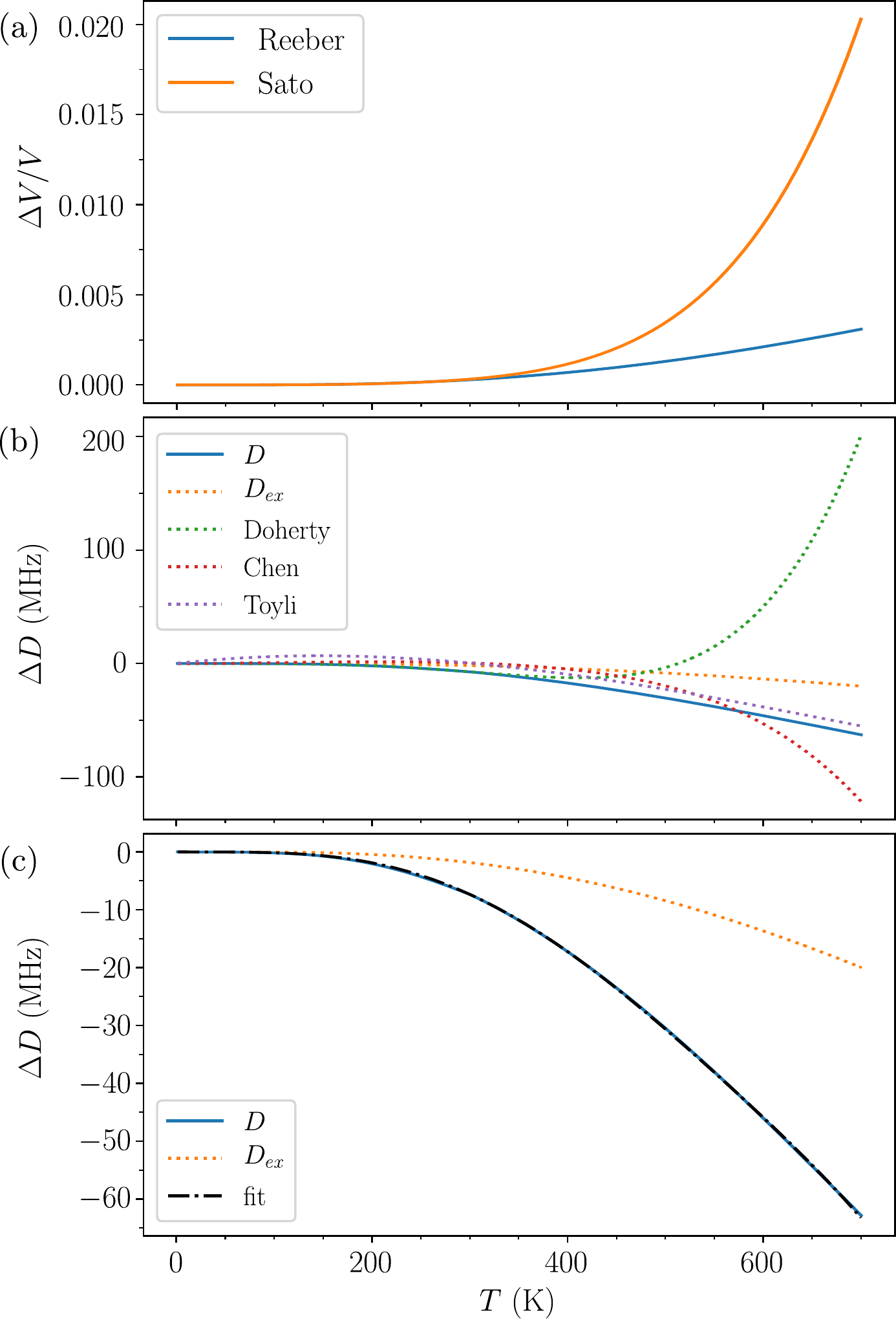}
	\caption{\textbf{(a)} Volumetric strain vs temperature using expressions from Sato \textit{et al} and Reeber \textit{et al}. Note the divergence for the description by Sato for $T > 300$ K. \textbf{(b)} The descriptions of the temperature dependence of $D(T)$ by Doherty \textit{et al}, Chen \textit{et al} and Toyli \textit{et al} with artificial data created by the expressions from Doherty and Toyli ($D$) and the expected result from considering thermal expansion strain only ($D_{ex}$). \textbf{(c)} Refitting over the full temperature range using equation (\ref{eqn-fit-bi}) and the artificial data ($D$).}
	\label{fig-appendix-DT}
\end{figure}

The multiple published expressions of the polynomial describing the temperature shift $\Delta D(T)$\cite{Acosta2010, PhysRevB.90.041201, chen2011temperature, toyli2012measurement} have been for narrower ranges in temperature presented here. Whilst these descriptions agree with each other within their prescribed temperature region they give wildly different results for predictions outside of their prescribed temperature range, as shown in figure \ref{fig-appendix-DT}(b). This divergence is a problem general to using a truncated polynomial or power series to describe a function. To circumvent this problem, we combined multiple expressions to obtain new polynomial parameters for the full useful temperature range of the \nv\, spin from liquid He temperature to 700 K. 

Doherty \textit{et al}\cite{PhysRevB.90.041201} described the temperature shift with a combination of the pressure shift due to the thermal expansion of the diamond ($a_i$ terms) and extra quadratic electron-phonon process ($b_i$ terms). The pressure shift is simply calculated from the hydrostatic stress parameter $A=14.6$ (MHz/GPa) \cite{doherty2014electronic}, the bulk modulus of diamond $B=442$ (GPa/strain) and the volumetric strain due to thermal expansion $\Delta V/V$. Doherty \textit{et al's} semi-empirical treatment resulted in suitable fit (for $T < 300$ K) using just two free parameters for the electron-phonon process $b_4$ and $b_5$.
\begin{align}
	\Delta D &= \Delta D_{ex} + \Delta D_{ep} \nonumber \\
	\Delta D &= -AB\frac{\Delta V}{V} - \sum_{i=4}^\infty b_i T^i \label{eqn-fit-bi} \\
	\Delta D &= -\sum_{i=2}^\infty a_i T^i - \sum_{i=4}^\infty b_i T^i \nonumber \\
	\begin{split}\Delta D &\approx\ - a_2 T^2 - a_3 T^3 - \\ &(b_4+a_4)T^4 - (b_5+a_5)T^5 + \cdots
	\end{split} \label{eqn-D(T)}
\end{align}

We adopt this method by refitting the $b_i$ terms to artificial $D$ data constructed from Toyli \textit{et al} \cite{toyli2012measurement} and Doherty \textit{et al}. This data was shifted to ensure a smooth curve, since we only fit the change in $D$ any change in offset is not important. However, Doherty \textit{et al's} model utilized the thermal expansion definition provided by Sato \textit{et al} \cite{PhysRevB.65.092102} which, for similar reasons, is only valid for temperatures up to 300 K, as shown in figure \ref{fig-appendix-DT}(a). As such, to further the temperature range we use the definition by Reeber \textit{et al} \cite{reeber1996thermal}, which is not provided in a convenient power series. To accurately describe the data we find it is necessary to fit up to the sixth power of $T$ and find the parameters $b_4 = -1.44(8)\times10^{-9}$, $b_5 = 3.1(3)\times10^{-12}$ and $b_6 = -1.8(3)\times10^{-15}$, this fit is shown in figure \ref{fig-appendix-DT}(c).

\section{Ensemble ODMR data analysis}
\label{section-appendix-ensemble-odmr-data}

\begin{figure}
	\centering
	\includegraphics[width=0.4\textwidth]{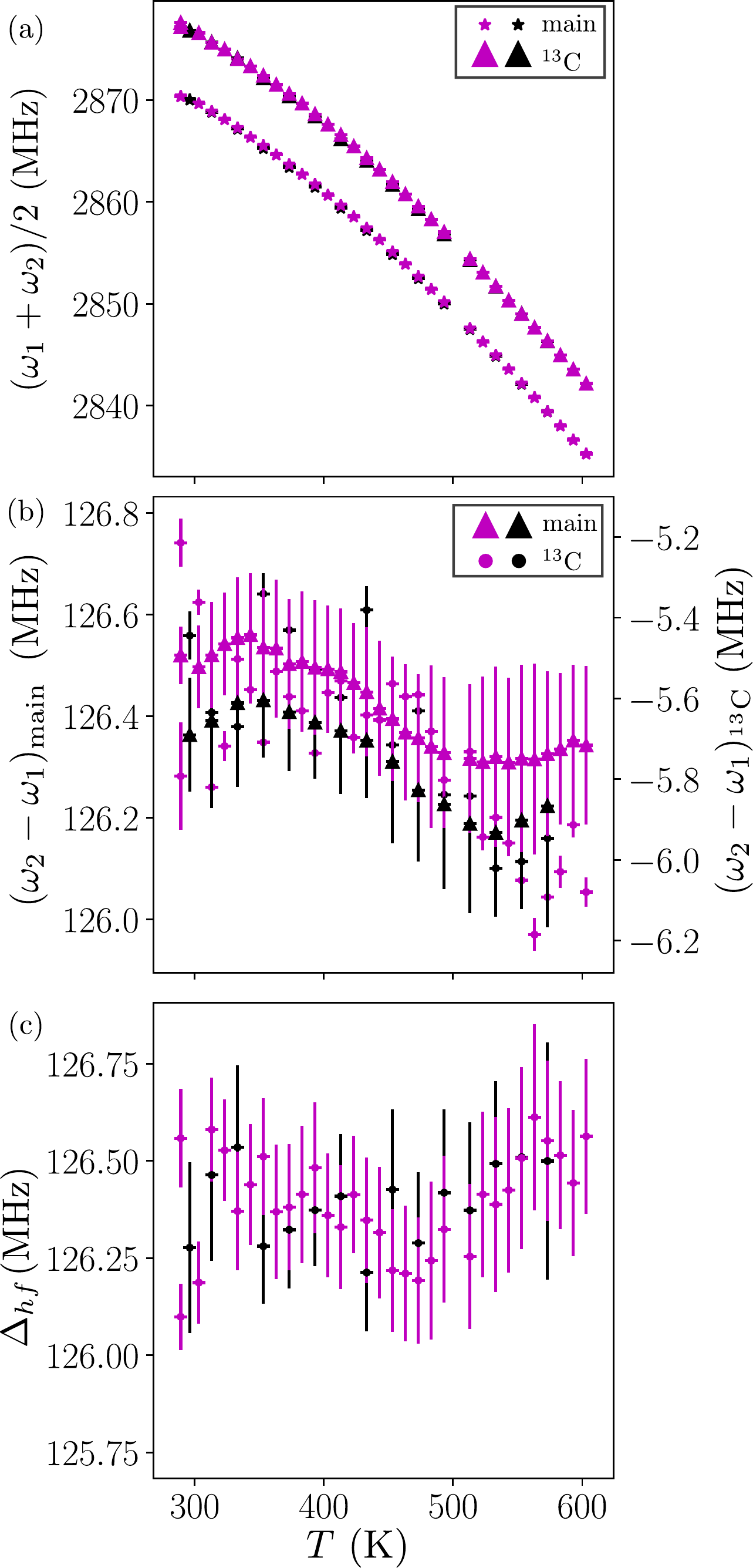}
	\caption{\textbf{(a)} The experimental data for the average spin-resonances from the high-temperature ensemble measurements for both the central main resonance and the \ct\, resonances. The different colors are from separate measurements. \textbf{(b)} The difference in spin-resonance of both the central main resonance and the \ct\, resonances. Note the non-monotonic behaviour of both spin-resonances with increasing temperature. This was repeatable for separate measurements. \textbf{(c)} The separation of the \ct\, spin-resonances with the temperature dependence of the separation of the main spin-resonances subtracted and shifted to match the low temperature data.}
	\label{fig-appendix-ensemble}
\end{figure}

The change in separation of the central ODMR spin-resonances (NV centers without a first shell \ct) demonstrated some unexpected behaviour with changing temperature. As shown in figure \ref{fig-appendix-ensemble}. This behaviour was observed to not be monotonic with increasing temperature and was repeatable across separate experimental runs. This was also present in the separation of the first shell \ct\,ODMR  spin-resonances. It was decided that this must be some remnant magnetization of the iron hotplate that has some thermal dependence, or some other non-linear process that effects the ODMR spectra. Another possibility is a change in strain or $^{14}$N hyperfine resonances, though how this occurs in a non-monotonic fashion is not clear. Since this behaviour was present in the both the \ct\, and non-hyperfine centers a like it was decided to simply subtract this effect from the separation of the \ct\, spin-resonances to separate the values of $\Delta_{hf}$.

\begin{thebibliography}{10}
	
	\bibitem{loubser1977optical}
	J.~H.~N. Loubser and J.~A. VanWyk.
	\newblock Optical spin-polarisation in a triplet state in irradiated and
	annealed type 1b diamonds.
	\newblock {\em Diamond Research}, pages 11--14, 1977.
	
	\bibitem{balasubramanian2008nanoscale}
	G.~Balasubramanian, I.~Y. Chan, R.~Kolesov, M.~Al-Hmoud, J.~Tisler, C.~Shin,
	C.~Kim, A.~Wojcik, P.~R. Hemmer, and A.~Krueger.
	\newblock Nanoscale imaging magnetometry with diamond spins under ambient
	conditions.
	\newblock {\em Nature}, 455(7213):648--651, 2008.
	
	\bibitem{maze2008nanoscale}
	J.~R. Maze, P.~L. Stanwix, J.~S. Hodges, S.~Hong, J.~M. Taylor, P.~Cappellaro,
	L.~Jiang, M.~V.~G. Dutt, E.~Togan, and A.~S. Zibrov.
	\newblock Nanoscale magnetic sensing with an individual electronic spin in
	diamond.
	\newblock {\em Nature}, 455(7213):644--647, 2008.
	
	\bibitem{taylor2008high}
	J.~M. Taylor, P.~Cappellaro, L.~Childress, L.~Jiang, D.~Budker, P.~R. Hemmer,
	A.~Yacoby, R.~Walsworth, and M.~D. Lukin.
	\newblock High-sensitivity diamond magnetometer with nanoscale resolution.
	\newblock {\em Nature Physics}, 4(10):810--816, 2008.
	
	\bibitem{dolde2011electric}
	F.~Dolde, H.~Fedder, M.~W. Doherty, T.~N{\"o}bauer, F.~Rempp,
	G.~Balasubramanian, T.~Wolf, F.~Reinhard, L.~C.~L. Hollenberg, F.~Jelezko,
	and J.~Wrachtrup.
	\newblock Electric-field sensing using single diamond spins.
	\newblock {\em Nature Physics}, 7(6):459--463, 2011.
	
	\bibitem{dolde2014charge}
	F.~Dolde, M.~W. Doherty, J.~Michl, I.~Jakobi, B.~Naydenov, S.~Pezzagna,
	J.~Meijer, P.~Neumann, F.~Jelezko, N.~B. Manson, and J.~Wrachtrup.
	\newblock Nanoscale detection of a single fundamental charge in ambient
	conditions using the $\mathrm{NV}{}^{\ensuremath{-}}$ center in diamond.
	\newblock {\em Physical Review Letters}, 112:097603, 2014.
	
	\bibitem{Acosta2010}
	V.~M. Acosta, E.~Bauch, M.~P. Ledbetter, A.~Waxman, L.~S. Bouchard, and
	D.~Budker.
	\newblock Temperature dependence of the nitrogen-vacancy magnetic resonance in
	diamond.
	\newblock {\em Physical Review Letters}, 104(7):70801, 2010.
	
	\bibitem{chen2011temperature}
	X.~Chen, C.~Dong, F.~Sun, C.~Zou, J.~Cui, Z.~Han, and G.~Guo.
	\newblock Temperature dependent energy level shifts of nitrogen-vacancy centers
	in diamond.
	\newblock {\em Applied Physics Letters}, 99(16):161903--161903, 2011.
	
	\bibitem{neumann2013high}
	P.~Neumann, I.~Jakobi, F.~Dolde, C.~Burk, R.~Reuter, G.~Waldherr, J.~Honert,
	T.~Wolf, A.~Brunner, and J.~H. Shim.
	\newblock High precision nano scale temperature sensing using single defects in
	diamond.
	\newblock {\em Nano Letters}, 2013.
	
	\bibitem{toyli2012measurement}
	D.~M. Toyli, D.~J. Christle, A.~Alkauskas, B.~B. Buckley, C.~G. {Van de Walle},
	and D.~D. Awschalom.
	\newblock Measurement and control of single nitrogen-vacancy center spins above
	600 k.
	\newblock {\em Physical Review X}, 2(3):031001, 2012.
	
	\bibitem{toyli2013fluorescence}
	D.~M. Toyli, F.~Charles, D.~J. Christle, V.~V. Dobrovitski, and D.~D.
	Awschalom.
	\newblock Fluorescence thermometry enhanced by the quantum coherence of single
	spins in diamond.
	\newblock {\em Proceedings of the National Academy of Sciences},
	110(21):8417--8421, 2013.
	
	\bibitem{nanomechanical_sensing_barson}
	M.~S.~J Barson, P.~Peddibhotla, P.~Ovartchaiyapong, K.~Ganesan, R.~L. Taylor,
	M.~Gebert, Z.~Mielens, B.~Koslowski, D.~A. Simpson, L.~P. McGuinness,
	J.~McCallum, S.~Prawer, S.~Onoda, T.~Ohshima, A.~C.~B. Jayich, F.~Jelezko,
	N.~B. Manson, and M.~W. Doherty.
	\newblock Nanomechanical sensing using spins in diamond.
	\newblock {\em Nano Letters}, 0(0):null, 2017.
	\newblock PMID: 28146361.
	
	\bibitem{teissier2014strain}
	J.~Teissier, A.~Barfuss, P.~Appel, E.~Neu, and P.~Maletinsky.
	\newblock Strain coupling of a nitrogen-vacancy center spin to a diamond
	mechanical oscillator.
	\newblock {\em Physical Review Letters}, 113(2):020503, 2014.
	
	\bibitem{ovartchaiyapong2014dynamic}
	P.~Ovartchaiyapong, K.~W. Lee, B.~A. Myers, and A.~C.~B. {Jayich}.
	\newblock Dynamic strain-mediated coupling of a single diamond spin to a
	mechanical resonator.
	\newblock {\em Nature Communications}, 5, 2014.
	
	\bibitem{PhysRevB.90.041201}
	M.~W. Doherty, V.~M. Acosta, A.~Jarmola, M.~S.~J. Barson, N.~B. Manson,
	D.~Budker, and L.~C.~L. Hollenberg.
	\newblock Temperature shifts of the resonances of the nv- center in diamond.
	\newblock {\em Physical Review B}, 90:041201, 2014.
	
	\bibitem{doherty2014electronic}
	M.~W. Doherty, V.~V. Struzhkin, D.~A. Simpson, L.~P. McGuinness, Y.~Meng,
	A.~Stacey, T.~J. Karle, R.~J. Hemley, N.~B. Manson, L.~C.~L Hollenberg, and
	S.~Prawer.
	\newblock Electronic properties and metrology applications of the diamond {NV-}
	center under pressure.
	\newblock {\em Physical Review Letters}, 112(4):047601, 2014.
	
	\bibitem{loubser1978electron}
	J.~H.~N. Loubser and J.~A. Wyk.
	\newblock Electron spin resonance in the study of diamond.
	\newblock {\em Reports on Progress in Physics}, 41:1201, 1978.
	
	\bibitem{doherty2012theory}
	M.~W. Doherty, F.~Dolde, H.~Fedder, F.~Jelezko, J.~Wrachtrup, N.~B. Manson, and
	L.~C.~L. Hollenberg.
	\newblock Theory of the ground-state spin of the {NV}$^-$ center in diamond.
	\newblock {\em Physical Review B}, 85(20):205203, 2012.
	
	\bibitem{felton2009hyperfine}
	S.~Felton, A.~M. Edmonds, M.~E. Newton, P.~M. Martineau, D.~Fisher, D.~J.
	Twitchen, and J.~M. Baker.
	\newblock Hyperfine interaction in the ground state of the negatively charged
	nitrogen vacancy center in diamond.
	\newblock {\em Physical Review B}, 79:075203, 2009.
	
	\bibitem{He1993paramagnetic-ii}
	X.~He, N.~B. Manson, and P.~T. Fisk.
	\newblock Paramagnetic resonance of photoexcited {NV} defects in diamond. {II}.
	hyperfine interaction with the $^{14}{N}$ nucleus.
	\newblock {\em Physical Review B}, 47(14):8816, 1993.
	
	\bibitem{ayscoughEPR1967}
	P.~B. Ayscough.
	\newblock {\em Electron Spin Resonance in Chemsitry}.
	\newblock Metheun \& Co Ltd, 1967.
	
	\bibitem{MORTON1978577}
	J.~R Morton and K.~F Preston.
	\newblock Atomic parameters for paramagnetic resonance data.
	\newblock {\em Journal of Magnetic Resonance (1969)}, 30(3):577 -- 582, 1978.
	
	\bibitem{vasp1996Kresse}
	G.~Kresse and J.~Furthm{\"u}ller.
	\newblock Efficient iterative schemes for ab initio total-energy calculations
	using a plane-wave basis set.
	\newblock {\em Phys. Rev. B}, 54:11169--11186, Oct 1996.
	
	\bibitem{PBEBurke1996}
	J.~P. Perdew, M.~Ernzerhof, and K.~Burke.
	\newblock Rationale for mixing exact exchange with density functional
	approximations.
	\newblock {\em The Journal of Chemical Physics}, 105(22):9982--9985, 1996.
	
	\bibitem{PhysRevB.65.092102}
	T.~Sato, K.~Ohashi, T.~Sudoh, K.~Haruna, and H.~Maeta.
	\newblock Thermal expansion of a high purity synthetic diamond single crystal
	at low temperatures.
	\newblock {\em Physical Review B}, 65:092102, 2002.
	
	\bibitem{reeber1996thermal}
	R.~R. Reeber and K.~Wang.
	\newblock Thermal expansion, molar volume and specific heat of diamond from 0
	to 3000k.
	\newblock {\em Journal of Electronic Materials}, 25(1):63--67, 1996.
	
	\bibitem{larsson2008electronic}
	J.~A. Larsson and P.~Delaney.
	\newblock Electronic structure of the nitrogen-vacancy center in diamond from
	first-principles theory.
	\newblock {\em Physical Review B}, 77(16):165201, 2008.
	
	\bibitem{manson1990raman}
	N.~B. Manson, X.~He, and P.~T. Fisk.
	\newblock Raman heterodyne detected electron-nuclear-double-resonance
	measurements of the nitrogen-vacancy center in diamond.
	\newblock {\em Optics Letters}, 15(19):1094--1096, 1990.
	
	\bibitem{Neumann1326}
	P.~Neumann, N.~Mizuochi, F.~Rempp, P.~Hemmer, H.~Watanabe, S.~Yamasaki,
	V.~Jacques, T.~Gaebel, F.~Jelezko, and J.~Wrachtrup.
	\newblock Multipartite entanglement among single spins in diamond.
	\newblock {\em Science}, 320(5881):1326--1329, 2008.
	
	\bibitem{dutt2007quantum}
	M.~V.~G. Dutt, L.~Childress, L.~Jiang, E.~Togan, J.~Maze, F.~Jelezko, A.~S.
	Zibrov, P.~R. Hemmer, and M.~D. Lukin.
	\newblock Quantum register based on individual electronic and nuclear spin
	qubits in diamond.
	\newblock {\em Science}, 316(5829):1312--1316, 2007.
	
\end{thebibliography}
\end{document}